\documentclass[
]{nrc1}                          

\usepackage{graphicx}       

\usepackage{amsmath}        

\usepackage{cite}           

\usepackage[french,english]{babel}

\volyear{XX}{2008}               
\journal{Can. J. Phys.}                       
\journalcode{cjp}                   
\filenumber{}                    

\received{}                      
\revreceived{}                   
\accepted{}                      
\revaccepted{}                   

\begin{document}



\title{The BMV project: Search for photon oscillations into massive particles}

\author{C. Robilliard}

\address[label1]{Laboratoire Collisions Agr\'{e}gats R\'{e}activit\'{e}, IRSAMC, UPS/CNRS,
UMR 5589, 31062 Toulouse, France.}

\correspond{email: cecile.robilliard@irsamc.ups-tlse.fr}

\author{B. Pinto Da Souza}
\address[label1]

\author{F. Bielsa}
\address[label2]{Laboratoire National des Champs Magn\'{e}tiques
Puls\'{e}s, CNRS/INSA/UPS, UMR 5147, 31400 Toulouse, France.}

\author{J. Mauchain}
\address[label2]

\author{M. Nardone}
\address[label2]

\author{G. Bailly}
\address[label1]

\author{M. Fouch\'e}
\address[label1]

\author{R. Battesti}
\address[label2]

\author{C. Rizzo}
\address[label1]

\shortauthor{C. Robilliard, B. Pinto Da Souza, F. Bielsa, {\it et al.}} 

\maketitle

\begin{abstract}
   In this contribution to PSAS08 we report on the research activities developed in
   our Toulouse group, in the framework of the BMV project, concerning the search for photon oscillations into massive
   particles, such as axion-like particles in the presence of a strong transverse
   magnetic field.
   We recall our main result obtained in collaboration with LULI at \'Ecole
   Polytechnique (Palaiseau, France). We also present the very
   preliminary results obtained with the BMV experiment which is set
   up at LNCMP (Toulouse, France).
\end{abstract}

\begin{resume}
Dans cette contribution \`a PSAS08, nous pr\'esentons les activit\'es de recherche d\'evelopp\'ees dans notre groupe \`a Toulouse, dans le cadre du projet BMV, concernant la recherche d'oscillations entre photons et particules massiques telles que les pseudo-axions en pr\'esence d'un champ magn\'etique transverse intense. Nous rappelons notre principal r\'esultat obtenu en collaboration avec le LULI \`a l'\'Ecole Polytechnique (Palaiseau, France). Nous pr\'esentons \'egalement les r\'esultats tr\`es pr\'eliminaires obtenus sur l'exp\'erience BMV, install\'ee au LNCMP (Toulouse, France).
\end{resume}


In Toulouse (France), since 2001 we have been conducting research
activities in the framework of the BMV (Bir\'efringence Magn\'etique
du Vide) project \cite{Askenazy2001} to study the propagation of
light in the presence of intense transverse magnetic fields, with
the goal to observe for the first time the vacuum magnetic
birefringence. This is a very fundamental prediction of quantum
electrodynamics that has never been experimentally proved (see
\cite{Battesti2008} and references therein).

In 1986, Maiani, Petronzio, and Zavattini \cite{Maiani} showed
that a hypothetical low mass, neutral, spinless boson, scalar or
pseudoscalar, that couples with two photons could induce an
ellipticity signal in the apparata designed to measure QED vacuum
magnetic birefringence \cite{Iacopini1979}. Moreover, an apparent
rotation of the polarization vector of light could be observed
because of the conversion of photons into real bosons, resulting in a
vacuum magnetic dichroism which is absent in the framework of
standard QED. The measurements of ellipticity and dichroism,
including their signs, can in principle completely characterize
the hypothetical boson, its mass $m_a$, the inverse coupling
constant $M$, and the pseudoscalar or scalar nature of the
particle. Ellipticity and dichroism are two different facets of
the same phenomenon: oscillation of photons into massive virtual
or real particles. Maiani, Petronzio, Zavattini's paper was
essentially motivated by the search for Peccei and Quinn's axion \cite{Peccei1977}.

The axion is a particle beyond the Standard Model. Proposed more than
30 years ago to solve the strong CP problem
\cite{Peccei1977,Weinberg_Wilczek}, this neutral, spinless,
pseudoscalar particle has not been detected yet, in spite of
constant experimental efforts
\cite{BNL1993,ADMX,CAST2007,Raffelt2007Review}. Whereas the most
sensitive experiments aim at detecting axions of solar or cosmic
origin, laboratory experiments including the axion source do not
depend on models of the incoming axion flux. Because the axion is
not coupled to a single photon but to a two-photon vertex,
axion-photon conversion requires an external electric or -
preferentially - magnetic field to provide for a virtual second
photon \cite{Sikivie1983}.

At present, purely terrestrial experiments are built according to
two main schemes. The first one, following the 1979 Iacopini and
Zavattini proposal \cite{Iacopini1979}, is sensitive to the
ellipticity and, slightly modified, to the dichroism induced by
the coupling of low mass, neutral, spinless bosons with laser beam
photons and the magnetic field \cite{Maiani}. The second popular
experimental scheme, named ``light shining through the wall''
\cite{VanBibber1987}, consists of first converting incoming
photons into axions in a transverse magnetic field, then blocking
the remaining photon beam with an opaque wall. Behind this wall
with which the axions do not interact, a second magnetic field
region allows the axions to convert back into photons with the
same frequency as the incoming ones. Counting these regenerated
photons, one can calculate the axion-photon coupling or put some
limits on it. This set-up was first realized by the BFRT
collaboration in 1993 \cite{BNL1993}.

Due to their impressive precision, optical experiments relying on
couplings between photons and these hidden-sector particles seem
most promising. Thanks to such couplings, the initial photons
oscillate into the massive particle to be detected. The strength
of optical experiments lies in the huge accessible dynamical
range: from more than $10^{20}$ incoming photons, one can be
sensitive to 1 regenerated photon!

In fact, the ``light shining through the wall'' experiment also
yields some valuable information on another hidden-sector
hypothetical particle \cite{Popov1991}. After the observation of a
deviation from blackbody curve in the cosmic background radiation
\cite{Woody1979}, some theoretical works suggested photon
oscillations into a low mass hidden sector particle as a possible
explanation \cite{Glashow1983}. The supporting model for such a
phenomenon is a modified version of electrodynamics proposed in
1982 \cite{Okun1982}, based on the existence of two U(1) gauge
bosons. One of the two can be taken as the usual massless photon,
while the second one corresponds to an additional massive particle
usually called paraphoton. Both gauge bosons are coupled, giving
rise to photon-paraphoton oscillations. Several years later, more
precise observations did not confirm any anomaly in the cosmic
background radiation spectrum \cite{COBERocket1990} and the
interest for paraphoton decreased, although its existence was not
excluded. More recently, it was found out that similar additional
U(1) gauges generally appear in string embeddings of the standard
model \cite{StringReviews}, reviving the interest for experimental
limits on the paraphoton parameters
\cite{Ahlers2007,Jaeckel2008,Ahlers2008}. Some limits on the mass
and the coupling constant of the paraphoton have already been
obtained by a photoregeneration experiment \cite{BNL1993}.

Motivated by the observation published by the PVLAS collaboration,
and subsequently retracted \cite{Zavattini2007}, which they
claimed could be explained by the existence of axions in the mass
range 1-2 meV, we have performed a ``light shining through the
wall'' experiment in collaboration with LULI at \'Ecole
Polytechnique (Palaiseau, France). We proved that the PVLAS signal
was not due to the existence of axions \cite{Robilliard2007}. We
recall here our main results \cite{Fouche2008}. We also present
the very preliminary results obtained with the BMV experiment
which is set up at LNCMP (Toulouse, France)
\cite{Battesti2008}. From the technological point of view, both
experiments rely on pulsed magnets to deliver the high
transverse magnetic fields needed. The coils we used have been
especially designed and developed in the framework of our project
\cite{Batut2008}.

\section{``Light shining through the
wall'' experiment at LULI}

This experiment has been described in details in ref.
\cite{Fouche2008}. We recall here the main points.

\subsection{Axion-like particles}

The photon to axion-like particle conversion and reconversion
transition probability (in natural units $\hbar=c=1$, with 1 T
$\equiv 195$\,eV$^2$ and 1~m $\equiv 5\times 10^6$\,eV$^{-1}$)
after propagating over a distance $L$ in a homogeneous magnetic
field $B_0$ writes

\begin{equation}
p_\mathrm{a} = \left(\Delta_M L \right)^2
\frac{\sin^2(\frac{\Delta_\mathrm{osc}}{2}L)}{(\frac{\Delta_\mathrm{osc}}{2}L)^2},
\label{eq:axion_p}
\end{equation}

\noindent where $\Delta_M = \frac{B_0}{2M} \quad \mbox{and}\quad
\Delta_\mathrm{osc} = \frac{m_\mathrm{a}^2}{2\omega}$, $\omega$
being the photon energy, $m_a$ the axion-like particle mass and
$M$ its inverse coupling constant with two photons. In our case,
our search was focused on the mass range $1 \, \mbox{meV} < m_a<2\,
\mbox{meV}$. Note that this equation is valid for a light
polarization parallel to the magnetic field since the axion has to
be a pseudoscalar \cite{Peccei1977}. Finally, for two identical
magnets, the photon regeneration probability due to axion-like
particles is $P_a=p_\mathrm{a}^2$.

\subsection{Paraphotons}

As far as paraphoton is concerned, in the modified version of
electrodynamics developed in 1982 \cite{Okun1982}, the paraphoton
weakly couples with the photon through kinetic mixing. Contrary to
axion-like particles, photon-paraphoton oscillations are therefore
possible without any external field and are independent on photon
polarization.

In the case of a typical photoregeneration experiment, the
incoming photons freely propagate for a distance $L_1$ and might
oscillate into paraphotons before being stopped by a wall, after
which the paraphotons propagate for a distance $L_2$ and have a
chance to oscillate back into photons that are detected. The photon regeneration
probability due to paraphotons can therefore be written as:

\begin{eqnarray}
P_{\gamma} & = & p_{\gamma}(L_1)p_{\gamma}(L_2) \nonumber \\
& = & 16\chi^4 \sin^2\left(\frac{\mu^2L_1}{4\omega}\right)
\sin^2\left(\frac{\mu^2L_2}{4\omega}\right)
\label{eq:paraphoton_regenerate}
\end{eqnarray}

\noindent where $\chi$ is the photon-paraphoton coupling constant,
and $\mu$ is the paraphoton mass which arbitrary values are to be
determined experimentally.

\noindent In our experiment, $L_1$ is the distance between the
focusing lens at the entrance of the vacuum system, which focuses
photons but not paraphotons, and the wall, which blocks photons
only. Similarly, $L_2$ represents the distance separating the
blind flange just before the regenerating magnet and the lens
coupling the renegerated photons into the optical fibre (see Fig.
\ref{fig:Setup}).

\subsection{Experimental set-up}

\begin{figure}[htb]
\begin{center}
  \includegraphics[width=8cm]{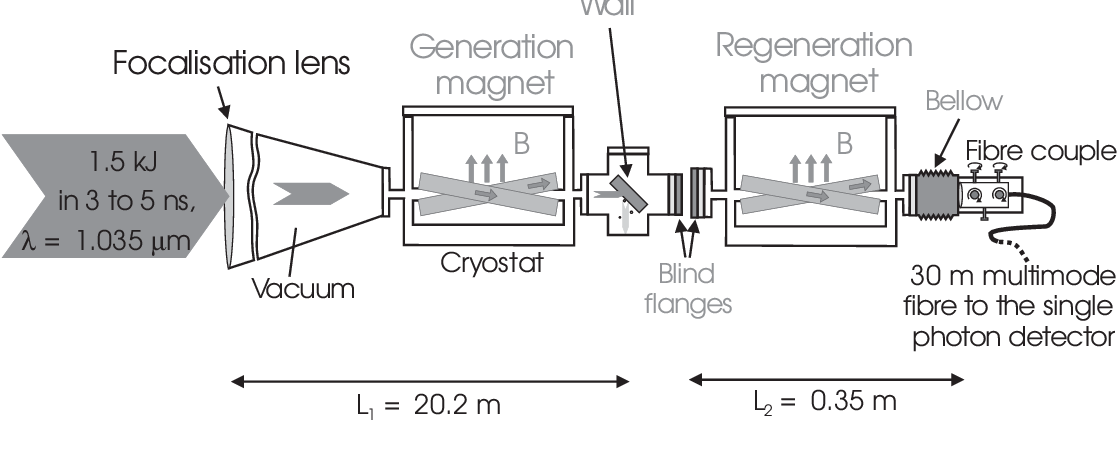}\\
  \caption{Sketch of the apparatus. The wall and the blind flanges
are removable for fibre alignment.}\label{fig:Setup}
\end{center}
\end{figure}

As shown in Fig.\,\ref{fig:Setup}, the experimental setup consists
of two main parts separated by the wall. An intense laser beam
travels through a first magnetic region (generation magnet) where
photons might be converted into axion-like particles. The wall
blocks every incident photons while axion-like particles would
cross it without interacting and may be converted back into
photons in a second magnetic region (regeneration magnet). The
regenerated photons are finally detected by a single photon
detector.

The three key elements leading to a high detection rate are the
laser, the generation and regeneration magnets placed on each side
of the wall and the single photon detector.

In order to have the maximum number of incident photons at a
wavelength that can be efficiently detected, the experiment has
been set up at Laboratoire pour l'Utilisation des Lasers Intenses
(LULI) in Palaiseau, on the Nano 2000 chain \cite{webLULI}. It can
deliver more than 1.5\,kJ over a few nanoseconds with $\omega =
1.17$\,eV. This corresponds to $N_i = 8\times10^{21}$\,photons per
pulse. During our 4 weeks of campaign, the total pulse duration ranged
from 5\,ns to 3\,ns while keeping the total energy constant.
At the end of the amplification chain, the vertically linearly
polarized incident beam has a 186\,mm diameter and is almost
perfectly collimated. It is then focused using a lens which focal
length is 20.4\,m. The wall is placed at $L_1 = 20.2$\,m from the
lens in order to have the focusing point a few centimeters behind
this wall. The beam is well apodized to prevent the incoming light
from generating a disturbing plasma on the sides of the vacuum
tubes.

In the region before the wall, where the laser beam propagates, a vacuum better
than $10^{-3}$\,mbar is necessary in order to avoid air
ionization. Two turbo pumps along the vacuum line easily give
$10^{-3}$\,mbar near the lens and better than $10^{-4}$\,mbar
close to the wall. The wall is made of a 15\,mm width aluminum
plate to stop every incident photon. It is tilted by 45\,$^\circ$
with respect to the laser beam axis in order to increase the area
of the laser impact and to avoid retroreflected photons. In the
second magnetic field region, a vacuum better than $10^{-3}$\,mbar
is also maintained.

Concerning the magnets, we use a pulsed technology. The pulsed
magnetic field is produced by a transportable generator developed
at LNCMP \cite{Frings2006}, which consists of a capacitor bank
releasing its energy in the coils in a few milliseconds. Besides,
a special coil geometry has been developed in order to reach the
highest and longest transverse magnetic field. The coil properties are
explained in details in Ref.\,\cite{Batut2008}. A coil consists of two
interlaced race-track shaped windings that are tilted one with
respect to the other. This makes room for the necessary optical
access at both ends in order to let the laser in while providing a
maximum $B_0 L$. Because of the particular
arrangement of wires, these magnets are called Xcoils.
The coil frame is made of G10 which is a non conducting material
commonly used in high stress and cryogenic temperature conditions.
External reinforcements with the same material have been added
after wiring to contain the magnetic pressure that can be as high
as 500\,MPa. A 12\,mm diameter aperture has been dug into the
magnets for the light path.
As for usual pulsed magnets, the coils are immersed in a liquid
nitrogen cryostat to limit the consequences of heating. The whole
cryostat is double-walled for a vacuum thermal insulation. This
vacuum is common with the vacuum line and is better than
$10^{-4}$\,mbar. A delay between two pulses is necessary for the
magnet to cool down to the equilibrium temperature which is
monitored via the Xcoils' resistance. Therefore, the repetition
rate is set to 5 pulses per hour.
The magnetic field is measured by a calibrated pick-up coil. The
maximum field $B_0$ is obtained at the center of the magnet.
Xcoils have provided $B_0 \geq 13.5$\,T over an equivalent length
$L = 365$\,mm. However, during the whole campaign a
lower magnetic field of $B_0=12\, (0.3)$\,T was used to increase
the coils' lifetime. The total duration of a magnetic pulse is a few milliseconds. The
magnetic field reaches its maximum value within less than 2\,ms
and remains constant ($\pm 0.3\%$) during $\tau_{B} = 150\,\mu$s,
a very long time compared to the laser pulse.

The last key element is the detector that has to meet several
criteria. In order to have as good a sensitivity as possible, the
regenerated photon detection has to be at the single photon level.
The integration time is limited by the longest duration of the
laser pulse which is 5\,ns. Since we expected about 100 laser
pulses during our four week campaign, which corresponds to a total
integration time of $500$\,ns, we required a detector with a dark
count rate far lower than 1 over this integration time, so that
any increment of the counting would be unambiguously associated to
the detection of one regenerated photon.
Our detector is a commercially available single photon receiver
from Princeton Lightwave which has a high detection efficiency at
$1.05\,\mu$m. It integrates a $80\times80\,\mu$m$^2$ InGaAs
Avalanche Photodiode (APD) with all the necessary bias, control
and counting electronics. Light is coupled to the photodiode
through a FC/PC connector and a multimode fiber. When the detector
is triggered, the APD bias voltage is raised above its reverse
breakdown voltage $V_{\mathrm{br}}$ to operate in ``Geiger mode''.
A short time later -- adjustable between 1\,ns and 5\,ns -- the
bias is reduced below $V_{\mathrm{br}}$ to avoid false events. For
our experiment, the bias pulse width is 5\,ns to correspond with
the longest laser pulse.
To optimize the dark count rate and the detection efficiency
$\eta_\mathrm{det}$, three different parameters can be adjusted:
the APD temperature, the discriminator threshold $V_\mathrm{d}$
set to reject electronic noise and the APD bias voltage
$V_{\mathrm{APD}}$. The dark count rate is first minimized by
choosing the lowest achievable temperature which is around 221\,K.
Dark counts for a 5\,ns detection gate as a function of
$V_\mathrm{d}$ increase rapidly when $V_\mathrm{d}$ is too low. On
the other hand, $\eta_\mathrm{det}$ remains constant for a large
range of $V_\mathrm{d}$. We set $V_\mathrm{d}$ to a value far from
the region where dark count increases and where
$\eta_\mathrm{det}$ is still constant. This corresponds to less
than $2.5\times10^{-2}$ dark count over $500$\,ns integration
time.
The detection efficiency is precisely measured by illuminating the
detector with a laser intensity lower than 0.1 photon per
detection gate at 1.05\,$\mu$m. The probability to have more than
one photon per gate is thus negligible. The best compromise
between detection efficiency and dark count rate is found for
$\eta_\mathrm{det} = 0.48 (0.025)$.\\

As said in the introduction, other similar experiments generally
require long integration times which implies an experimental
limitation due to the detection noise. Using pulsed laser,
magnetic field and detection is an original and efficient way to
overcome this problem. Photons are concentrated in very intense
short laser pulses during which the detection background is
negligible. This also means that if a photon is detected in our
experiment in correlation with the magnetic field, it will be an
unambiguous signature of axion generation inside our apparatus.\\

After the second magnet, the regenerated photons are injected into
the detector through a coupling lens and a graded index multimode
fiber with a 62.5\,$\mu$m core diameter, a 0.27 numerical aperture
and an attenuation lower than 1\,dB/km. These parameters ensure
that we can inject light into the fiber with a high coupling
ratio, even when one takes into account the pulse by pulse
instability of the propagation axis that can be up to 9\,$\mu$rad.
During data acquisition, the mean coupling efficiency through the
fibre was found to be $\eta_c = 0.85$; it was optimized just before each pulse with an attenuated beam originating from the same pilot, hence exactly superimposed to the high energy beam used for the measurements.
The only remaining source of misalignment lies in thermal effects
during the high energy pulse, which could slightly deviate the
laser beam, hence generating supplementary losses in fibre
coupling. This misalignment is mostly reproducible and can be
corrected by acting on the mirrors steering the beam to the wall.

Our experiment is based on pulsed elements which require a perfect
synchronization : the laser pulse must cross the magnets when the
magnetic field is maximum and fall on the photodiode during the
detection gate.
The magnetic pulse is triggered with a TTL signal from the laser
chain. The magnetic trigger has a jitter lower than 10\,$\mu$s,
ensuring that the laser pulse travels through the magnets within
the 150\,$\mu$s interval during which the magnetic field is
constant and maximum.
Synchronization of the laser pulse and the detector needs to be
far more accurate since both have a 5\,ns duration. The detector
gate is triggered with the same fast signal as the laser, using
delay lines. We have measured the coincidence rate between the
arrival of photons from the attenuated beam described above on the detector and the opening of the 5\,ns
detector gate as a function of an adjustable delay \cite{Fouche2008}. We have chosen
our working point in order to maximize the coincidence rate.

\subsection{Data analysis}

The best experimental limits are achieved if no fake signal is
detected during the experiment, which was indeed the case. To estimate the
corresponding upper conversion probability of regenerated photons,
we have to calculate the upper number of photons that could have
been missed by the detector $n_\mathrm{missed}$ for a given
confidence level ($CL$), which writes

\begin{equation}
n_\mathrm{missed} =
\frac{\log(1-CL)}{\log(1-\eta_\mathrm{det})}-1.
\label{eq:n_missed}
\end{equation}

\noindent For example, with our value of $\eta_\mathrm{det}$, a
confidence level of 99.7\,$\%$ corresponds to less than 8 missed
photons. The upper photon regeneration probability is then
\begin{equation}
P_{\mathrm{a}\;\mathrm{or}\;\gamma} =
\frac{n_\mathrm{missed}}{N_{\mathrm{eff}}}, \label{eq:Proba}
\end{equation}
where $N_{\mathrm{eff}}$ is the number of effective incident
photons over the total number of laser shots, taking into account
the losses described before.

Data acquisition was spread over 4 different weeks. 82 high energy
pulses have reached the wall with a total energy of about 110\,kJ.
This corresponds to $5.9\times10^{23}$ photons. During the whole
data acquisition, no signal has been detected.
The magnetic field was applied during 56 of those laser pulses,
with a mean value of 12\,T. The laser pulses without magnetic
field aimed at testing for possible fake counts.

\begin{figure}[htb]
\begin{center}
\includegraphics[width=8cm]{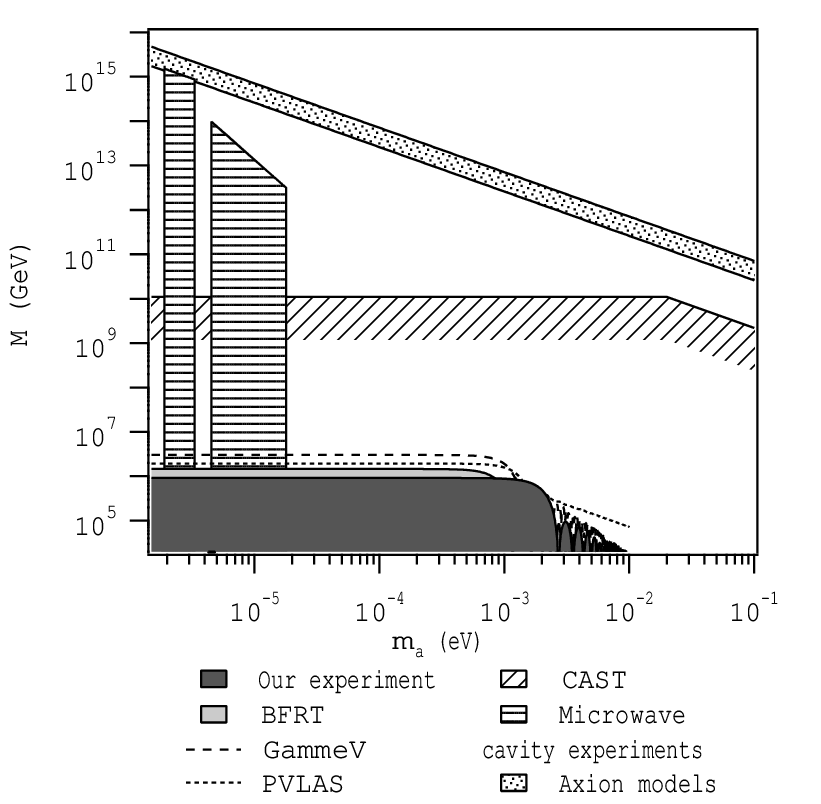}
\caption{Limits on the axion-like particle - two photon inverse
coupling constant $M$ as a function of the axion-like particle
mass $m_\mathrm{a}$ obtained by experimental searches. Our
exclusion region is first compared to other purely laboratory
experiments such as the BFRT photon regeneration experiment
\cite{BNL1993}, the GammeV experiment \cite{FermiLab2008} and the
PVLAS collaboration \cite{PVLAS2008} with a 3$\sigma$ confidence
level. Those curves are finally compared to the 95\,$\%$
confidence level exclusion region obtained on CAST \cite{CAST2007}
and the more than 90\,$\%$ confidence level on microwave cavity
experiments \cite{CavityRBF,CavityUF,ADMX}. Model predictions are
also shown as a dotted stripe between the predictions of the KSVZ
model (lower line, $E/N=0$) \cite{KSVZ} and of the DFSZ model
(upper line, $E/N=8/3$) \cite{DFSZ}.}
\label{fig:CourbeGenerale_Axion}
\end{center}
\end{figure}

Our experimental sensitivity limits for axion-like particle at
$99.7\,\%$ confidence level correspond to a detection probability
of regenerated photons $P_\mathrm{a} = 3.3\times10^{-23}$ and give
$M>9.1\times10^5$\,GeV at low masses \cite{Fouche2008}.
We show our limits together with those from other laboratory experiments in
Fig.\,\ref{fig:CourbeGenerale_Axion}. They are comparable to the limits obtained by other
purely laboratory experiments
\cite{BNL1993,FermiLab2008,PVLAS2008}, especially in the meV
region of mass. On the other hand, they are still much less sensitive than experiments which limits (stripes) approach model predictions
\cite{CavityRBF,CavityUF,ADMX,CAST2007}.

\begin{figure}[htb]
\begin{center}
\includegraphics[width=8cm]{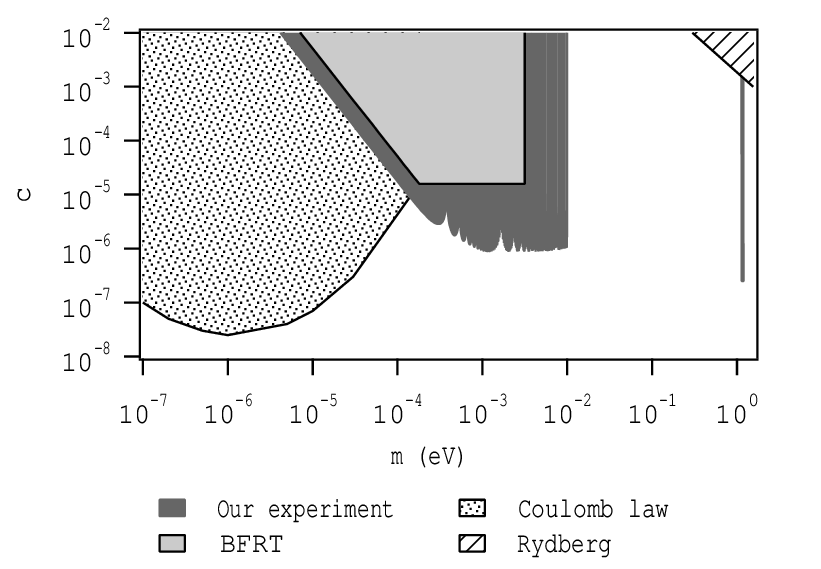}
\caption{$95\,\%$ confidence level limits on photon-paraphoton
mixing parameter as a function of the paraphoton mass obtained thanks to
our null result (deep gray area). Shaded regions are excluded. This
is compared to excluded regions obtained on BFRT photon regeneration
experiment \cite{BNL1993} (light gray area), to searches for
deviations of the Coulomb law \cite{CoulombLaw} (points) and to
comparisons of the Rydberg constant for different atomic transitions
\cite{Rydberg} (stripes).} \label{fig:CourbeGenerale_Paraphoton}
\end{center}
\end{figure}

In the case of paraphotons our measurements correspond to a
maximum photon regeneration probability $P_\gamma =
9.4\times10^{-24}$. This sets a limit $\chi < 1.1\times10^{-6}$
for 1\,meV$<\mu<$10\,meV with a $95\,\%$ confidence level. As shown in Fig. \ref{fig:CourbeGenerale_Paraphoton}, this
improves by about one order of magnitude the exclusion area
obtained on BFRT photon regeneration experiment \cite{BNL1993}.
Comparing to other laboratory experiments
\cite{CoulombLaw,Rydberg} (see \cite{Review_ParaExp} for review),
we were able to constrain the paraphoton parameters in a region
which had not been covered so far by purely terrestrial
experiments.

\section{The BMV experiment at LNCMP}

This experiment has been described in details in ref.
\cite{Battesti2008}. We recall here the main points and present
our very preliminary results.

Linearly polarized light, propagating in a medium in the presence
of a transverse magnetic field, acquires an ellipticity
\cite{RizzoRizzo}. Indeed, the velocity of light propagating in the
presence of a transverse magnetic field B depends on the light
polarization, i.e. the index of refraction $n_\parallel$
for light polarized parallel to the magnetic field is different
from the index of refraction $n_\perp$ for light polarized
perpendicular to the magnetic field. For symmetry reasons, the
difference  $\Delta n = (n_\parallel - n_\perp)$ is proportional
to $B^2$. Thus, in general an incident linearly polarized light
beam exits elliptically polarized from the magnetic field region.
Quantum ElectroDynamics (QED) predicts that a field of 1 T should
induce an anisotropy of the index of refraction of vacuum $\Delta
n$ of about $4\times10^{-24}$ \cite{Bialynicka-Birula,Adler}.

Photon oscillations into a virtual massive particle like axions
also induce an ellipticity signal $\psi$ in such an apparatus
\cite{Maiani}. This ellipticity can be written as \cite{BNL1993} :

\begin{equation}
\psi = \left(\frac{\Delta_M^2 L}{\Delta_\mathrm{osc}} \right)
\left(1 -
\frac{\sin(\Delta_\mathrm{osc})}{\Delta_\mathrm{osc}}\right),
\label{eq:axion_ell}
\end{equation}

\noindent where $\Delta_M = \frac{B_0}{2M} \quad \mbox{and}\quad
\Delta_\mathrm{osc} = \frac{m_\mathrm{a}^2}{2\omega}$, $\omega$
being the photon energy, $m_a$ the axion-like particle mass and
$M$ its inverse coupling constant with two photons. Moreover, since we use an optical resonant cavity to increase the
optical path in the magnetic field region, the ellipticity
acquired by the light is increased by a factor $2 {F}\over{\pi}$,
where $F$ is the finesse of the optical cavity. The finesse is
related to the photon lifetime $\tau$ in the cavity by the
formula $\tau$ = $F L\over{\pi c}$, where $c$ is the light velocity. 

\begin{figure}[h!]
\begin{center}
\includegraphics[width=9cm]{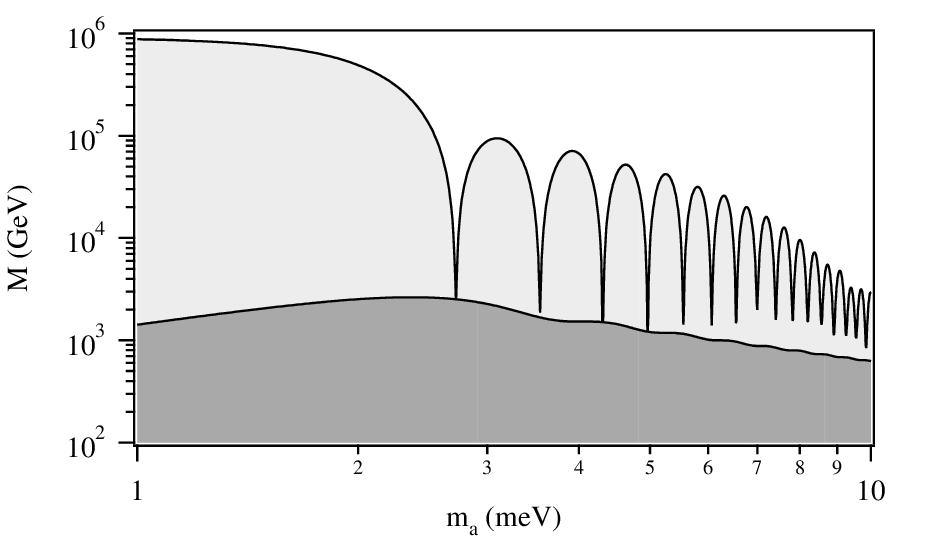}
\caption{Excluded mass $m_a$ and inverse coupling constant $M$ for axion-like particles, obtained respectively by the measurement of a zero ellipticity on the BMV experiment (dark grey) and the null result of our photoregeneration experiment at LULI (light grey).} \label{Limits}
\end{center}
\end{figure}

During our first run once the whole apparatus was operational, we have
measured the magnetic birefringence of molecular Nitrogen. Our
experimental value $\Delta n =(-2.49\pm0.05)\times10^{-13}$ at 1
atm and 273.15 K is in good agreement with other existing values
\cite{RizzoRizzo}. We have also acquired data in vacuum to
search for vacuum magnetic birefringence. After 17 magnetic field
pulses with $B_0\simeq 9$ T over a length about 0.5 m and a cavity
finesse $F\simeq 3000$, we have reached a value $\Delta n$ per T$^2$ of
$\left(-10\pm23\right)\times10^{-17}$ T$^{-2}$, which is obviously compatible
with zero. In Fig. \ref{Limits} we show the limit obtained on the parameters of axion-like particles by
this ellipticity measurement, together with the one obtained by our
group at LULI.

\begin{figure}[h!]
\begin{center}
\includegraphics[width=8cm]{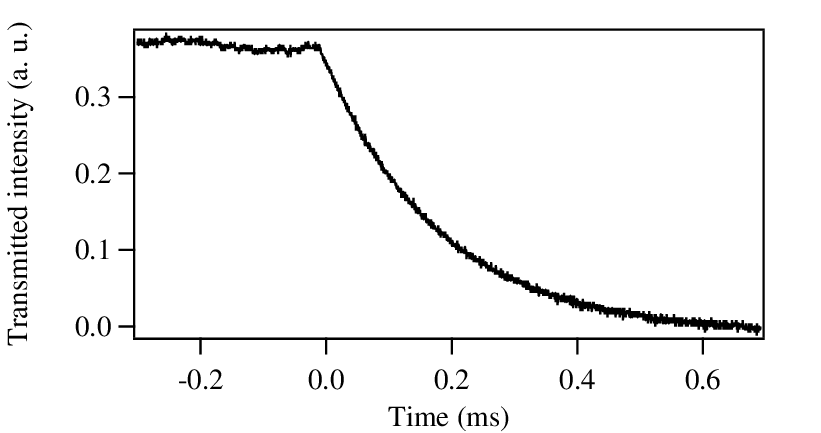}
\caption{Typical measurement of photon lifetime in the
optical cavity, corresponding to $\tau \simeq 190 \,\mu$s and $F\simeq 80000$.} \label{Finesse}
\end{center}
\end{figure}

Recently we have upgraded our optical cavity to reach a higher finesse and thus a better sensitivity. In Fig. \ref{Finesse} we show a photon lifetime measurement,
corresponding to $\tau\simeq 190 \,\mu$s and
$F\simeq 80000$\cite{mirrors}. Such a photon lifetime in a cavity
is one of the longest ever measured (see e.g. \cite{Deriva}).

\section{Conclusion and Outlooks}

We have presented the final results of our photon regeneration
experiment which excluded the PVLAS results. Our null measurement
leads to limits similar to other purely terrestrial axion
searches, and improves the preceding limits by more than one order
of magnitude concerning paraphotons \cite{Ahlers2007}. 

Some further improvements on axion-like particles are to be expected from our BMV experiment, which is now operational. A first run has been performed.
The limit obtained by ellipticity measurements is still far from
being innovative but it is nevertheless encouraging, and the final set-up should allow us to improve by one to two orders of magntitude the present best limit on axion-like particles from purely terrestrial experiments.

More generally, let us argue that such precision optical
experiments may prove useful for experimentally testing the
numerous theories beyond standard model in the low energy window,
a range in which the large particle accelerators are totally
helpless. For example, our apparatus can be modified to become
sensitive to chameleon fields \cite{Chameleon}.


\begin{thebibliography}{99}
\bibitem{mirrors} Mirrors have been manufactured by Layertec company.
\end{thebibliography}
\end{document}